\documentclass[aps,prl,twocolumn]{revtex4-1}
\usepackage[latin1]{inputenc}
\usepackage{amsmath}
\usepackage{setspace}
\usepackage[margin=0.5in]{geometry}
\usepackage{xcolor}
\usepackage{grffile}
\usepackage{hyperref}
\usepackage{float}
\usepackage{csquotes}
\usepackage{amsfonts}
\usepackage{amssymb}
\usepackage{graphicx}
\usepackage{float, tikz}
\definecolor{green}{RGB}{34,139,34}
\begin{document}

\newcommand\plotSize{0.485\linewidth}
\title{Resilience of networks of multi-stable chaotic systems to targetted attacks }
\author{ Chandrakala Meena$^{1,2}$, Pranay Deep Rungta$^2$
\thanks{e-mail: meenachandrakala@gmail.com} and Sudeshna Sinha$^2$\thanks{e-mail: sudeshna@iisermohali.ac.in}}
\address{$^1$ Department of Mathematics, Bar-Ilan University, Ramat Gan, 5290002, Israel,\\
$^2$ Department of Physical Sciences, Indian Institute of Science Education and Research Mohali, SAS Nagar, Sector 81, PO Manauli 140306, Punjab, India}

\begin{abstract}
  We investigate the collective dynamics of chaotic multi-stable Duffing oscillators connected in different network topologies, ranging from star and ring networks, to scale-free networks. We estimate the resilience of such networks by introducing a variant of the concept of multi-node Basin Stability, which allows us to gauge the global stability of the collective dynamics of the network in response to large perturbations localized on certain nodes.
        We observe that in a star network, perturbing just the hub node has the capacity to destroy the collective state of the entire system. On the other hand, even when a majority of the peripheral nodes are strongly perturbed, the hub manages to restore the system to its original state. This demonstrates the drastic effect of the centrality of the perturbed node on the collective dynamics of the full network. Further, we explore scale-free networks of such multi-stable oscillators and demonstrate that targetted attacks on nodes with high centrality can destroy the collective dynamics much more efficiently than random attacks, irrespective of the nature of the nodal dynamics and type of perturbation. We also find clear evidence that the betweeness centrality of the perturbed node is most crucial for dynamical robustness, with the entire system being more vulnerable to attacks on nodes with high betweeness. These results are crucial for deciding which nodes to stringently safeguard in order to ensure the recovery of the network after targetted localized attacks.
\end{abstract}

\maketitle
Spatiotemporal patterns emerging in the collective dynamics of complex systems are determined by the interplay of the dynamics of each node and the nature of the interactions among the nodes \cite{baruch2013,uzi2017,cnsns}.
Interactions among nodes are often modelled by a gamut of coupling topologies and coupling forms, and these interactions influence spatiotemporal pattern formation in the network \cite{hens2019,meena2016,nir2018,sync,extreme,symmetry,explosive}.
A question of utmost relevance in dynamical networks is the ability of the system to recover from large perturbations. In this context it is crucial to ascertain how significantly the properties of an individual node impact the collective dynamics of a network. The influence of a node on the global dynamics is expected to depend on how critical the node is to information pathways in the network. Deciphering this correlation will enable us to identify the nodes which render the network most susceptible to targetted perturbations. This in turn will help determine the nodes to safeguard most stringently from external influences in order to maintain the collective state.

In this work we focus of the following nodal features: 

(i) Degree of a node $i$, denoted by $k_i$, defined as the number of neighbors that are directly connected to the node,
in an undirected network. High degree of a node indicates that the node is in direct contact with a larger set of nodes. 

(ii) Normalized betweeness centrality of a node $i$ \cite{chen,pre_hong}, defined as:
$$b_i= \frac{2}{(N-1)(N-2)} \sum_{s,t\in \mathcal{N}}\frac{\sigma(s,t|i)}{\sigma(s,t)}$$ 
where $\mathcal{N}$ is the set of all nodes, $\sigma(s,t)$ is the number of shortest paths between nodes $s$ and $t$ and $\sigma(s,t|i)$ is the number of shortest paths passing through the node $i$. If node $i$ has high betweeness centrality, it implies that it lies on many shortest paths, and so there is a high probability that the communication pathway between any two nodes passes through it.

(iii) Normalized Closeness Centrality, defined as:
$$c_i=\frac{N-1}{\sum_{j} d(j,i)}$$
where $d(j,i)$ is the shortest path between node $i$ and node $j$ in the graph. It is the inverse of the average length of the shortest path between the node and all other nodes in the network\cite{alex}. Consequently high closeness centrality implies short communication paths to different parts of the network, as the number of steps to reach the other nodes is small.

Since the above features of a node determine the efficiency of the transfer of information emanating from it, or through it, they are expected to influence the propagation of perturbations originating at the node. 

Now, an important feature of nonlinear dynamical systems is {\em multi-stability}, namely the co-existence of multiple attractors, of varying degrees of dynamical complexity, for a given set of parameters \cite{pisarchik2014control}.  Networks of such multi-stable systems have been receiving much research attention in recent times and is important in many areas of natural sciences, such as neuronal science \cite{schiff, foss}, ecosystem \cite{may1977,groffman2006ecological}, condensed matter physics \cite{Prengel,bonilla2006voltage}, optics \cite{prasad_2003,Brambilla_1991}, environmental science \cite{chandrakala_csf, chandrakala_pramana}, chemistry \cite{calistus,hudson1981chaos}, biology \cite{ullner2008multistability,koseska2007inherent}.
Due to switching among co-existing attractors, multi-stable systems are characterized by a high degree of complexity in dynamical behaviour, and are typically very sensitive to initial conditions and perturbations. Some attractors may be advantageous in specific engineered applications or natural situations, while it may be detrimental being trapped in others. So if we want to target or maintain the robustness of a particular attractor in a complex system \cite{control,computation_with_multistability,threshold_2017}, understanding the stability of the entire system in the presence of strong perturbations plays a very important role \cite{mitra}.
 
Synchronization in networks of coupled dynamical systems with two co-existing fixed points has been investigated recently \cite{our_pre}, focussing on the stability of the collective state where all elements are in the vicinity of the same fixed point attractor.
In this work we go beyond such simple co-existing attractors, and focus on the dynamics of multi-stable networks supporting complex attractors \cite{expt, Lind,kapitaniak}.
Specifically we will consider networks of multi-stable Duffing oscillators, with co-existing limit cycles and chaotic attractors, as a generic test-bed to explore the robustness of complex multi-stable networks.

\begin{figure}[htb]
	
	\centering
	\includegraphics[width=0.95\linewidth]{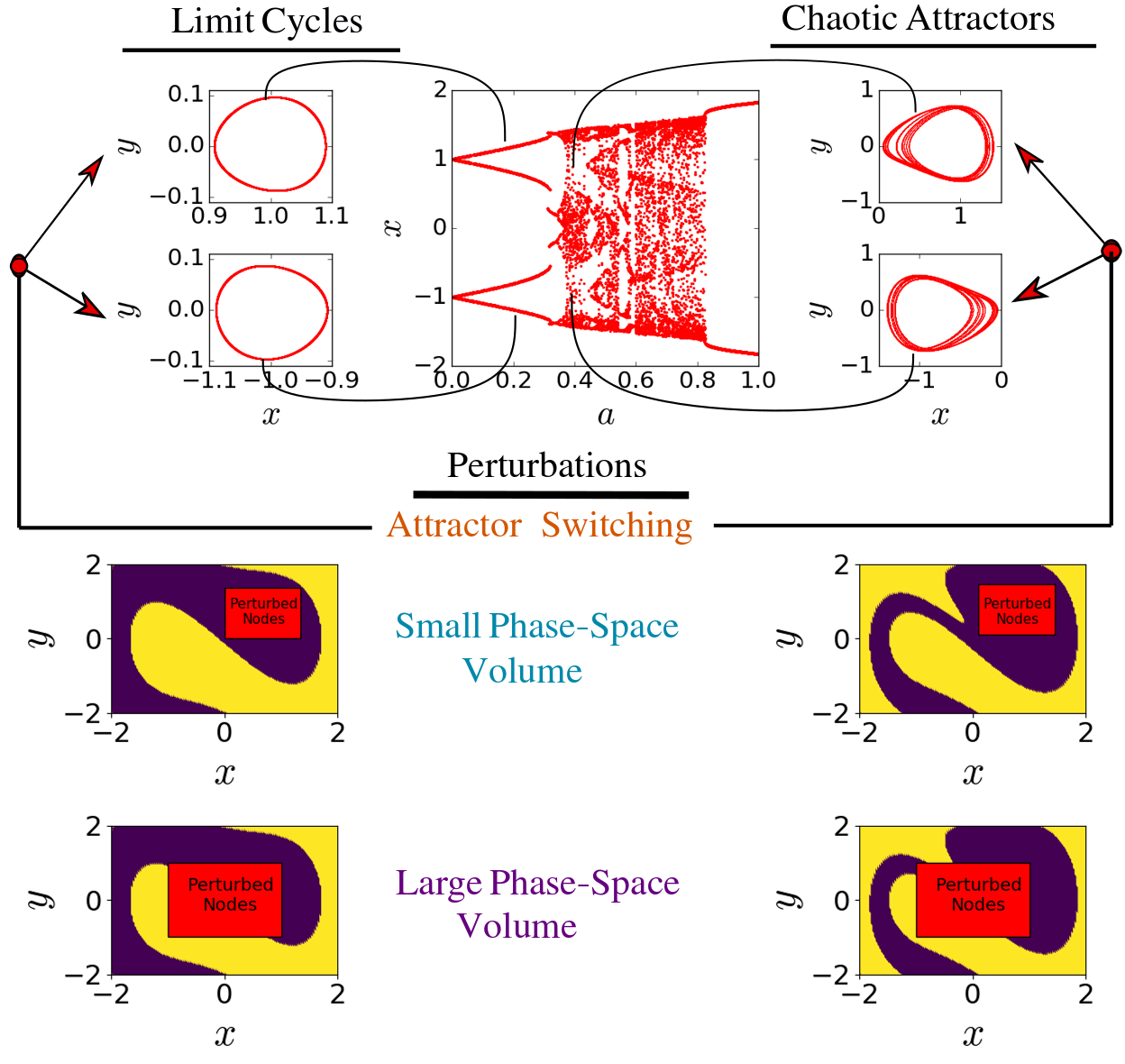}
	
	\caption{Bifurcation diagram of the state of the Duffing oscillator given by Eqn. \ref{unit} (with the $x$-variable displayed), as a function of amplitude of periodic forcing $a$, showing co-existing limit cycles and chaotic attractors. Basins of attraction of the co-existing attractors in phase space are also shown in the figure. The yellow and purple colors indicate the attractor which has positive and negative values of the state variable $x$ respectively. The red box represents the phase space volume where the state of the nodes are perturbed onto, under different perturbation schemes. Note that when the nodes are perturbed into a small volume of phase space, the perturbation sends the state into the basin of attraction of the other attractor, while perturbations onto a large volume of phase space spans both basins of attraction. }
	\label{main_fig}
\end{figure}

The Duffing oscillator is governed by a set of two first order nonautonomous differential equations given by:
\begin{eqnarray}
\label{unit}
\dot{x}&=&f_x (x, y)= y\\ \nonumber
\dot{y}&=&f_y (x, y)= F(x) - \delta y + a \sin (\omega t) 
\end{eqnarray}
where $F(x) = x - x^{3}$, $a$ is the amplitude, $\omega$ is the angular frequency of the periodic driving force and $\delta$ controls the amount of damping. Eqn. \ref{unit} describes an externally driven particle in a two-well non-parabolic potential. By adjusting any of the control parameters $\delta, a$ or $\omega$, we can observe multi-stable attracting states that are dynamically more complex, such as co-existing limit cycles and chaotic attractors (cf. Fig. \ref{main_fig}). Here we consider parameter values $\delta=0.5$, $\omega=1$, and we vary the amplitude of the periodic driving force given by parameter $a$. Duffing oscillators have been used extensively to study anharmonic and chaotic oscillations, and serves as a standard model of nonlinear dynamics \cite{chaos_duffing}, as despite its simplicity the emergent dynamical behaviour is extremely rich. It particular, it has been successfully used to model a variety of physical processes such as stiffening springs \cite{spring}, beam buckling \cite{beam}, non-linear electronic circuits \cite{circuit}, and ionization waves in plasmas \cite{plasma}. 

First we analyse the dynamical behaviour of the Duffing oscillators and identify the parameter values where co-existing attractors are present. We look for both coexisting limit cycles, as well as coexisting chaotic attractors. It is evident from the bifurcation diagram (cf. Fig. \ref{main_fig}) that the Duffing oscillator displays rich dynamics of co-existing limit cycles and chaotic attractors, bounded in distinct regions of phase space. In particular, we find that there is multi-stable limit cycles in the parameter range $0<a<0.35$ and multi-stable chaotic attractors in the range $0.35<a<0.85$. The figure also displays the basins of attraction for the case of co-existing limit cycles (at $a=0.1$), as well as co-existing chaotic attractors (at $a=0.36$). It is clear that the basin boundary is more complex for co-existing chaotic attractors.
 
Now we will go on to explore the dynamics of such multi-stable Duffing oscillators diffusively coupled in different network topologies. The general form of a network of size $N$, comprised of local nodal dynamics and coupling interactions, can be given as:

\begin{eqnarray}
\label{main}
\dot{x_i}&=& f_x (x_i, y_i) + C \frac{1}{k_i} \sum_j A_{ij} (x_j - x_i)\\ \nonumber
&=& f_x (x_i, y_i) + C (\langle x_i^{nn} \rangle - x_i)\\ \nonumber
\dot{y_i}&=& f_y (x_i, y_i) 
\end{eqnarray}

where index $i$  ($i = 1, \dots N$) specifies a node in the network, and the nodal dynamics $(f_x, f_y)$is given by the Duffing oscillator (cf. Eqn.~\ref{unit}). The parameter $C$ determines the strength of coupling and $k_i$ is the degree of node $i$. The connectivity matrix $A_{ij}$ reflects the topology of the underlying connections. Here we consider rings, star networks and scale-free networks \cite{scalefree,cohen2010}. This form of coupling implies that each nonlinear oscillator evolves under the influence of a \enquote{local mean field} generated by the coupling neighbourhood of each site $i$, $\langle x_i^{nn} \rangle = \frac{1}{k_i} \sum_j x_j$, where $j$ is the node index of the neighbours of the $i^{th}$ node and {\em `nn'} indicates the nearest neighbours of a node $i$. In order to investigate the case of co-existing limit cycles, we consider the value of parameter $a=0.1$ in the nodal dynamics given by Eqn.~\ref{unit}, and for the case of co-existing chaotic attractors we consider $a=0.36$.

Our focus will be on the robustness of the collective state, where all nodes in the network are in the basin of the same attractor,
and we will attempt to correlate nodal properties, such as degree, betweeness and closeness centrality, with the ability of the network to revert to the original state, in response to large perturbations on specific nodes.
To gauge the global stability and robustness of a state, we will {\em use a variant of the recent multi-node Basin Stability formalism} \cite{menck2013basin,mitra,our_pre}. In general, the Basin Stability (BS) of a particular attractor of a multi-stable dynamical system is given by the fraction states that return to the original basin of the attraction after large perturbation. We estimate the multi-node Basin Stability using the follwing method. At the outset we will consider all nodes of the network localized on one of the two stable attractors, i.e. $\{ x_i, y_i \}$ of all nodes $i$ lie on different phase-space points of one of the attractors. We then give a large perturbation to some fraction $f$ of nodes.

We consider three distinct types of perturbations (c.f. Fig. \ref{main_fig}) in this work:

(i) Attractor Switching:  here we perturb the nodes on to randomly chosen phase-space points lying on a different attractor, i.e. the attractors are switched for the perturbed nodes. For instance, for the case of Duffing oscillators with two co-existing limit cycles, if the initial states of the nodes in the network are localized on the attractor with negative $x$ values, then the perturbed nodes will lie on the limit cycle with positive $x$ values. 

(ii) Small Phase-Space Volume Perturbation: Here the perturbations send the state of the oscillator at the perturbed node onto a small phase-space volume in the basin of attraction of another co-existing attractor.

(iii) Large Phase-Space Volume Perturbation: Here the perturbations send the state of the oscillator at the perturbed node onto a phase point that is randomly located in a large volume of phase-space, and this volume typically includes sections from different basins of attraction. This type of perturbation has been most commonly considered in previous studies. For instance, for the case of coupled Duffing oscillators we perturb the nodes onto a randomly chosen phase-space point in the phase-space box $x \in [-1:1]$ and  $y \in [-1:1]$.

Note that when the nodes are perturbed into a small volume of phase space, the perturbation always sends the state of the perturbed node into the basin of attraction of the other attractor, while perturbations onto a large volume of phase space have some probability of being in either basin of attraction.

After perturbations, we ascertain whether all the oscillators return to their original attractors, i.e. we find out whether or not the perturbed system recovers completely to the state it was initially in. We repeat this ``experiment'' over a large sample of perturbation strengths and sets of perturbed nodes, and find the fraction of instances the system manages to revert to the original state. This measure of global stability is then a variant of multi-node Basin Stability and reflects the robustness of the collective state to perturbations localized at particular nodes in the network.
The power of this concept arises from the fact that it determines the probability of the system to remain in the basin of attraction of a particular state when random perturbations affect a specific number and type of nodes. This enables one to extract the contributions of individual nodes to the global stability of the collective dynamics of the complex network. So this variant of multi-node Basin Stability can help discern the {\em nodal characteristics that make the network more vulnerable to targetted attack}, as one can perturb subsets of nodes with certain specified nodal features.

In general then, we will attempt to identify the {\em characteristics of the perturbed nodes that most significantly influence the robustness of the multi-stable dynamical network.} That is, we will search for discernable patterns in nodal features that help sustain the collective network dynamics, as well as those that destroy it. Specifically, we will try to correlate the centrality features of the perturbed nodes, with the recovery of a network with complex co-existing attractors from large localized perturbations. 

In the following sections,
we will first consider rings, where all nodes have the same centrality. This provides us with a reference system to understand how the fraction of perturbed nodes influences resilience. We will then go on to study the star network, where the degree, closeness and betweeness centrality of the central hub node is very different from that of the peripheral nodes. This network then provides a clean system to investigate the correlation between the centrality of a node and the resilience of the network under large localized perturbations at such nodes. Lastly we will consider scale-free networks, which are often encountered in the real world. This heterogeneous network will enable us to answer a more subtle question, namely which centrality property (degree, closeness or betweeness) of a perturbed node is most crucial in determining the potential recovery of the network.

\section{Ring of multi-stable oscillators}

We first consider a ring of Duffing oscillators where the initial states of all the nodes are distributed over phase points of one of the attractors, i.e. the constituent oscillators inhabit the same region of phase space, associated with the basin of one, or the other, of the coexisting dynamical states. We then strongly perturb a fraction $f$ of nodes such that the states of the perturbed nodes are kicked to phase points of the other coexisting attractor i.e. the attractors are switched for the perturbed nodes. We examine the case where the perturbed nodes are in clusters, i.e. the perturbed nodes are contiguous to each other, and also the case where the perturbed nodes are randomly distributed over the ring.

\begin{figure}[htb]
	\centering
	\includegraphics[width=1\linewidth]{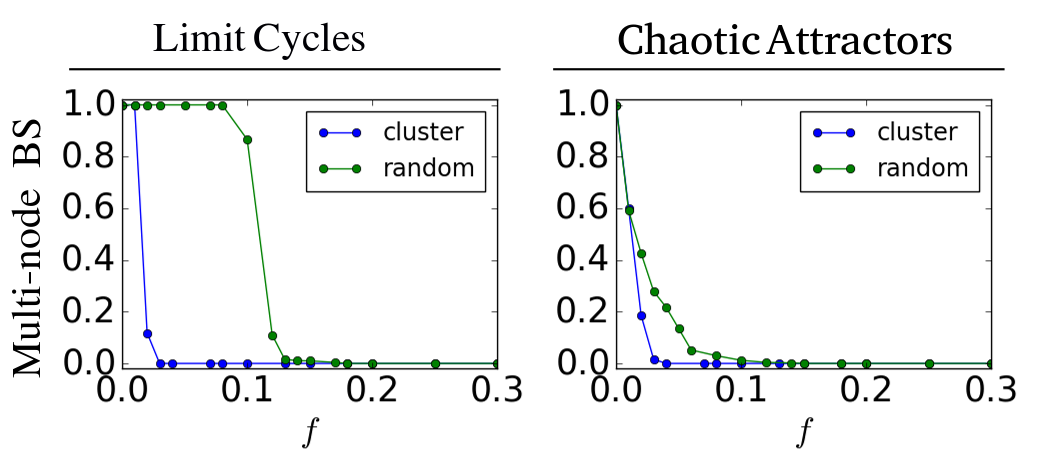}
	\caption{Dependence of multi-node Basin Stability on fraction $f$ of nodes perturbed, in a ring of Duffing oscillators given by Eqn.~\ref{main}, with $N=100$, $C=1$ and $k=2$ (i.e. each site couples to its two nearest neighbours), for multi-stable limit cycles (left panel) and multi-stable chaotic attractors (right panel). The green curve represents the case where the perturbed nodes are chosen at random locations and the blue curve shows the case of perturbations in a spatial clusters.}
	\label{bsvsc_ring}
\end{figure}

Fig.~\ref{bsvsc_ring} demonstrates the sensitive dependence of the multi-node Basin Stability on fraction $f$ of perturbed nodes. For instance, when a localized cluster of $3$ nodes is perturbed in a ring of $100$ multi-stable limit cycles or multi-stable chaotic attractors (i.e. $f=0.03$), the Basin Stability drops to zero, implying that the probability of the system to return to the original state is negligible. Hence a ring rapidly loses its ability to recover from perturbations with increasing fraction of perturbed nodes. In general, it is also evident that a ring of multi-stable limit cycles is more robust than a network of multi-stable chaotic attractors.

From Fig.~\ref{bsvsc_ring} also demonstrates another interesting feature. The capacity of the ring to recover from perturbations depends significantly on whether the perturbed nodes are spatially located in a cluster (cf. blue curves) or randomly located over the ring (cf. green curves). We observe that the system is more stable when random nodes are targetted for perturbations, as compared to the case when nodes are perturbed in clusters. This effect is more pronounced for multi-stable limit cycles, which retain resilience upto significantly larger number of perturbed nodes for the case of random perturbation locations. So one can infer that {\em perturbations at random locations in a ring allows the system to recover its original dynamical with more ease than perturbations on a cluster of contiguous nodes}. 

\section{Star network of multi-stable oscillators}

Next we study a star network of multi-stable Duffing oscillators. This is an interesting test bed, as there are two very distinct classes of nodes: the single central hub node and the peripheral nodes surrounding it, and the degree, closeness and betweeness centrality of the hub is much larger than that of all the other nodes at the periphery. Therefore, this network provides a good framework to investigate the correlation between centrality of a node and the resilience of the network in the presence of large targetted attacks on such nodes.

In this study we consider a star network where the initial states of all the nodes are distributed over phase points of one of the attractors. We then give a large perturbation to a fraction $f$ of nodes, pushing  the phase space points of the perturbed nodes onto the other coexisting attractor (i.e. attractor switching scheme), and estimate the multi-node Basin Stability of the network. 
\begin{figure}[htb]
	\centering
	\includegraphics[width=1\linewidth]{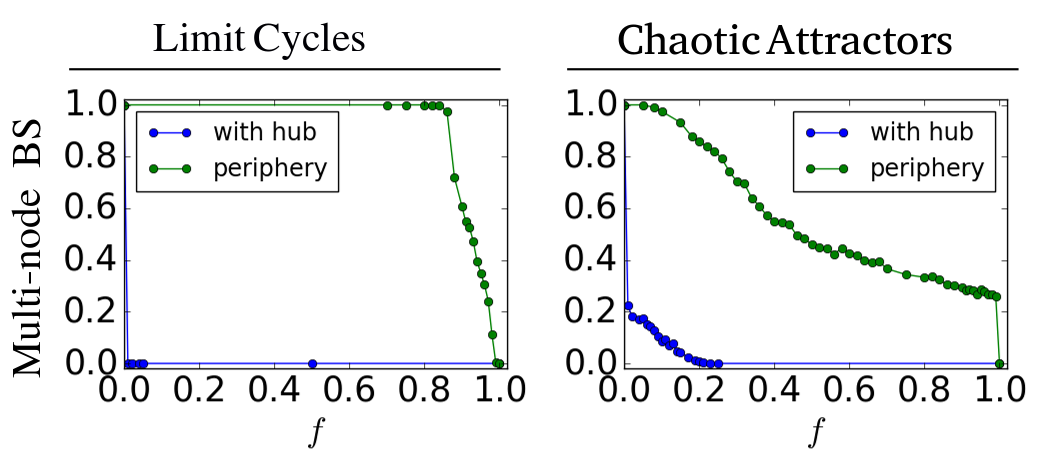}
	\caption{Dependence of multi-node Basin Stability on fraction $f$ of nodes perturbed, in the star network of Duffing oscillators, with $N=100$ and $C=1$. The blue curve represents the case where the perturbed nodes always include the hub node, while the green curve represents the case where peripheral nodes are perturbed in varying fractions and the hub node is perturbed only when $f=1$. In the left panel $a=0.1$ (i.e. the case of multi-stable limit cycles) and in the right panel $a=0.36$ (i.e.  the case of multi-stable chaotic attractors).}
	\label{bsvsc_star}
\end{figure}

We obtain the dependence of the multi-node Basin Stability on fraction $f$ of perturbed nodes, ranging from single node to the case where nearly all nodes in the system are perturbed (cf. Fig.~\ref{bsvsc_star}). We examine two distinct situations. In the first case the set of perturbed nodes always includes the hub, shown in Fig. \ref{bsvsc_star} by the blue curve. In the second situation, the peripheral nodes are perturbed, and only for the case of $f=1$ (i.e. when all nodes perturbed) is the hub also perturbed. This case is shown by the green curve in the figure. It is evident from Fig. \ref{bsvsc_star} (left panel), that for networks of Duffing oscillators with multi-stable limit cycles, even for $f$ is as high as $0.8$ the Basin Stability remains close to $1$, i.e. even when  $80 \%$ of the peripheral nodes are strongly perturbed the entire network almost always recovers to the original state. So the hub node is capable of steering the large number of perturbed peripheral nodes back to their original state.  For the case of networks of Duffing oscillators with multi-stable chaotic attractors however, perturbations on a few peripheral nodes (e.g. $f \sim 0.1$) can cause the network to start losing global stability (cf. Fig. \ref{bsvsc_star} right panel). So the degree of resilience of a star network to localized perturbations on low centrality peripheral nodes depends on the type of nodal dynamics, with chaotic attractors being more vulnerable to rapid destabilization.

Focussing now on the case where the perturbed nodes always include the high centrality hub node (shown in blue curves) we observe that the Basin Stability of a network of Duffing oscillators with multi-stable limit cycles, as well as multi-stable chaotic attractors, drops steeply to very low values when $f$ becomes non-zero. So one can infer that {\em just a single perturbed node of very high centrality, such as the hub, can destroy the ability of the network to recover its initial state, irrespective of the nature of the nodal dynamics.} 

\section{Scale-Free network of multi-stable oscillators}

Lastly we investigate scale-free networks of multi-stable Duffing oscillators, constructed via the Barab{\'a}si-Albert preferential attachment algorithm, with parameter $m$ reflecting the number of links attached to each new node \cite{scalefree}. Since this is a heterogeneous network, it offers a valuable test-bed for studying the effects of perturbations at nodes with a range of centrality properties. Further, we can exploit the very marked difference in the distributions of degree, betweeness and closeness centrality of the nodes in scale-free networks with $m=1$ vis-a-vis $m=2$, to probe which particular centrality measure most affects network robustness.

\begin{figure}[htb]
	\centering
	\includegraphics[width=0.98\linewidth]{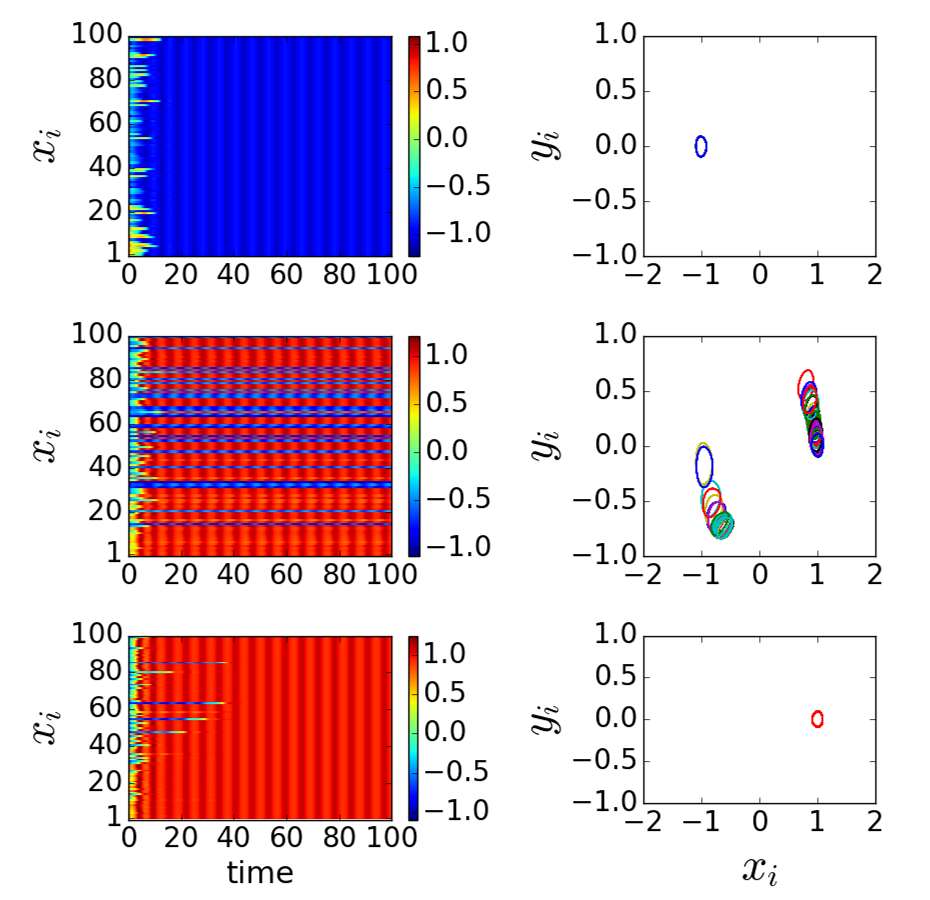}
	\caption{Time evolution and phase portraits of $100$ Duffing oscillators coupled in a scale-free network with $m=2$, given by Eqn.~\ref{main}, with coupling strength $C = 1$ and $a=0.1$, under the attractor switching perturbation scheme. Here the perturbed nodes are of highest betweeness centrality, and the fraction of perturbed nodes is (a) $0.15$, (b) $0.25$, (c) $0.3$ and (d) $0.35$.}
	\label{RSFspt_hbtclimit}
\end{figure}

We first study the influence of the fraction $f$ of perturbed nodes on the collective network dynamics, through space-time plots and phase space plots. With no loss of generality, we show representative results for networks of multi-stable limit cycles of size $N=100$, with $a=0.1$. Fig. \ref{RSFspt_hbtclimit} shows the emergent dynamics when nodes with the highest betweeness centrality are perturbed, with the initial states of the nodes randomly distributed on the attractor with negative $x$. It is evident that when the fraction of perturbed nodes is relatively small, the perturbed nodes return to their original states and move back to their initial attractor (cf. top row). When $f$ is moderately large, the perturbed nodes remain in the attractor onto which they have been perturbed, and also induce some of the other nodes to switch states (cf. middle row). When the fraction of perturbed nodes reach a critical value $f_{crit}$, then {\em all} oscillators are dragged away from their original attractor, i.e. the oscillators at all the nodes switch  attractors (cf. bottom row). So as the fraction of perturbed nodes increases, there is a transition from a collective state where all oscillators are in one particular attractor, to a collective state where all oscillators move to the other attractor. For instance in Fig.~\ref{RSFspt_hbtclimit}, the network moves from the state where all oscillators have negative $x$, to a state where all have positive $x$, and this transition occurs via a short range where these two states co-exist.

Now we will investigate two important questions. First, we would like to explore how different properties of the perturbed nodes (degree, betweeness and closeness centrality) affect the resilience of the collective state of the multi-stable network. Secondly, we would like to ascertain how the complexity of the dynamical attractors affects dynamical robustness. 

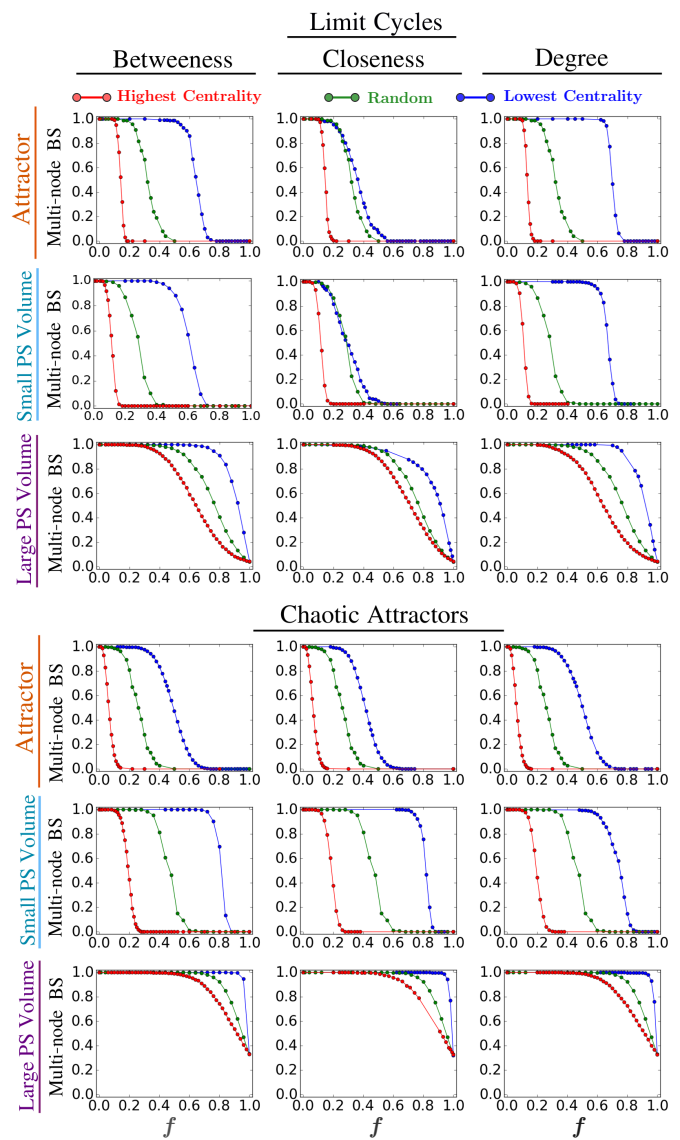
\begin{figure}[h]
\begin{tikzpicture}
\node (img) at (-1.5,0)	
{\includegraphics[width=0.98\linewidth]{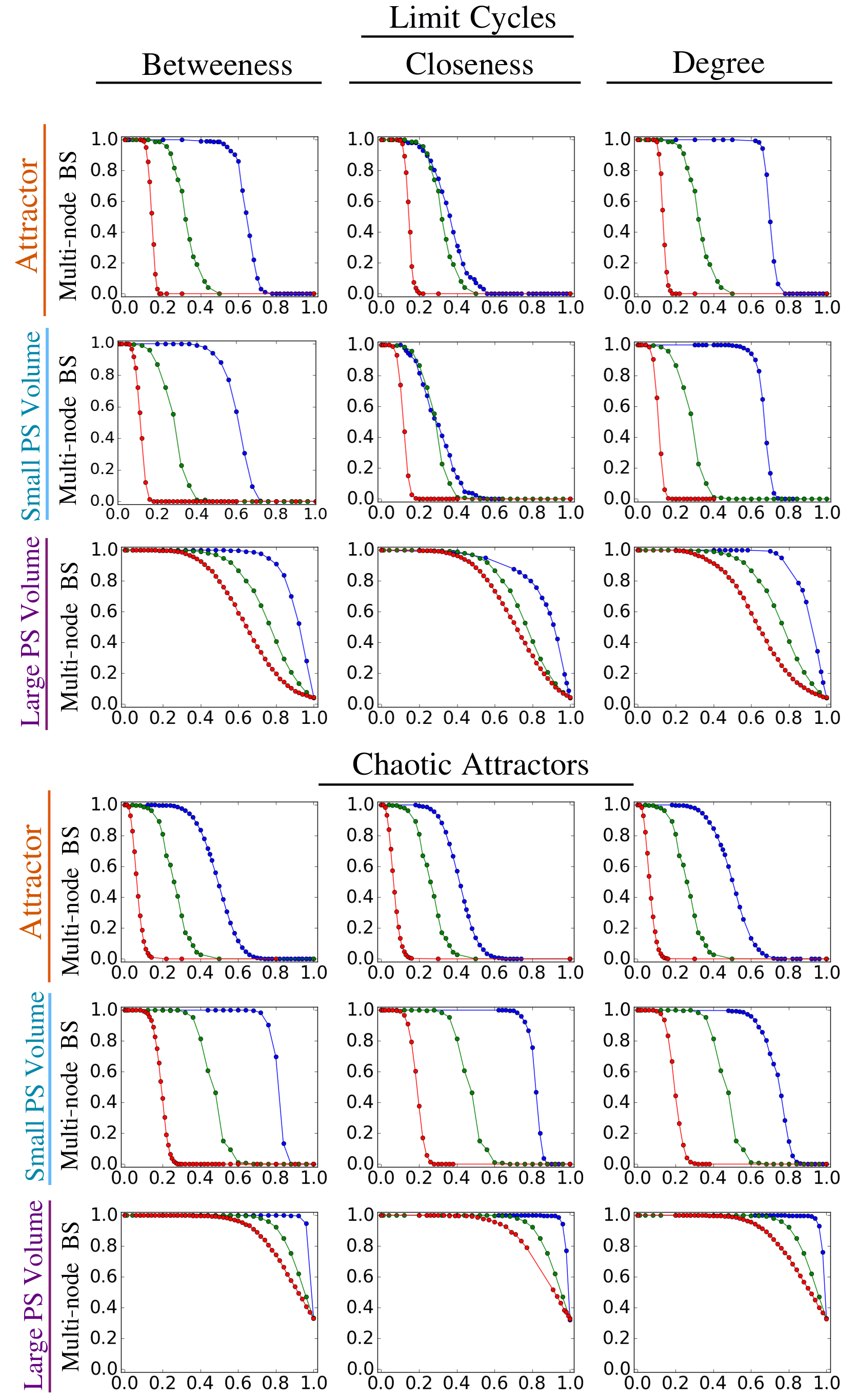}};
\node[scale=0.6, rotate=0,left of= g] (specJ) at (-3.15,-7.5) {\bf \textcolor{black!70}{\Large \textit{f}}};
\node[scale=0.6, rotate=0,left of= g] (specJ) at (-0.4,-7.5) {\bf \textcolor{black!80}{\Large \textit{f}}};
\node[scale=0.6, rotate=0,left of= g] (specJ) at (2.3,-7.5) {\bf \textcolor{black!90}{\Large \textit{f}}};
\draw [red!100, thick] (-5.0,6.2) -- (-4.6, 6.2) node[] {}
;
\filldraw[color=black!80, fill=red!60] (-5.0,6.2) circle (1.5pt) node[anchor=west]{};
\filldraw[color=black!80, fill=red!60] (-4.6,6.2) circle (1.5pt) node[anchor=west]{};
\node[scale=0.5, rotate=0,left of= g] (specJ) at (-3.0,6.2) {\bf \textcolor{red!100}{\large Highest Centrality}};

\draw [green!100, thick] (-1.65,6.2) -- (-1.3, 6.2) node[] {}
;
\filldraw[color=black!80, fill=green!80] (-1.65,6.2) circle (1.5pt) node[anchor=west]{};
\filldraw[color=black!80, fill=green!80] (-1.3,6.2) circle (1.5pt) node[anchor=west]{};

\node[scale=0.5, rotate=0,left of= g] (specJ) at (-0.2,6.2) {\bf \textcolor{green!100}{\large Random}};

\draw [blue!100, thick] (0.45,6.2) -- (0.1, 6.2) node[] {}
;
\filldraw[color=black!80, fill=blue!80] (0.1,6.2) circle (1.5pt) node[anchor=west]{};
\filldraw[color=black!80, fill=blue!80] (0.5,6.2) circle (1.5pt) node[anchor=west]{};
\node[scale=0.5, rotate=0,left of= g] (specJ) at (2.1,6.2) {\bf \textcolor{blue!100}{\large Lowest Centrality}};
\end{tikzpicture}
\caption{Dependence of the multi-node Basin Stability of scale-free networks of multi-stable oscillators, on the fraction $f$ of perturbed nodes, with $N=100$, $m=2$, $C=1$, and $a=0.1$ (multi-stable limit cycles) and $a=0.36$ (multi-stable chaotic attractors), for three cases: (green curves) perturbed nodes chosen at random; (red curves) perturbed nodes chosen in descending order of betweeness centrality (left panels), closeness centrality (middle panels) and degree (right panels), i.e, where the perturbed nodes have the highest $b$, $c$ or $k$; (blue curves) perturbed nodes chosen in ascending order of betweeness centrality (left panels), closeness centrality (middle panels) and degree (right panels), i.e. the perturbed nodes have the lowest $b$, $c$ or $k$. Attractor switching, small phase-space volume and large phase-space volume perturbation schemes are considered (cf. Fig.~\ref{main_fig}). The behaviour is not very sensitive to system size.
}
\label{fnode_properties}
\end{figure}	

In order to systematically explore the correlation between nodal properties and resilience, we rank the nodes in increasing (or decreasing) order of the different centrality measures we are probing (such as degree, closeness or betweeness). We then go on to examine the behaviour of the network when certain fraction $f$ of nodes with the highest (or lowest) centrality are strongly perturbed. That is, we estimate the multi-node Basin Stability under large perturbations on sub-sets of nodes having the highest (or lowest) centrality.
It is clearly evident again from Fig.~\ref{fnode_properties}, as is intuitively expected, that {\em when the perturbed nodes have the highest degree, closeness and betweeness centrality, there is drastic reduction of Basin Stability, while the Basin Stability falls significantly slower when nodes of low centrality are perturbed.}
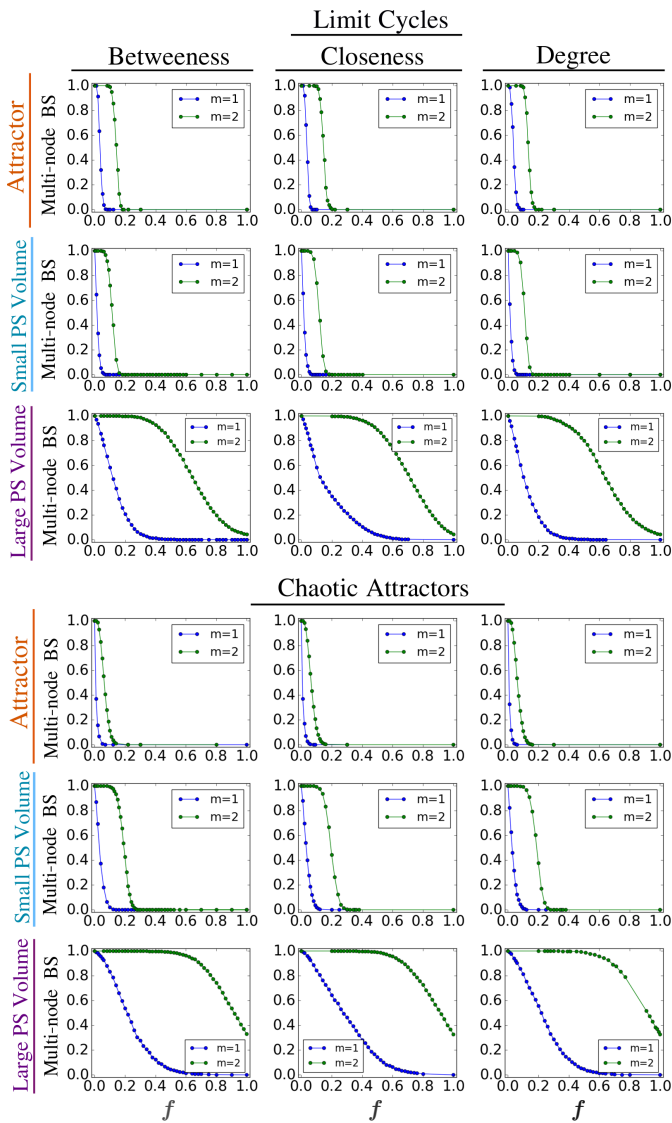
\begin{figure}[htb]
	\centering
	\begin{tikzpicture}
	\node (img) at (-1.5,0)	
	{\includegraphics[width=0.98\linewidth]{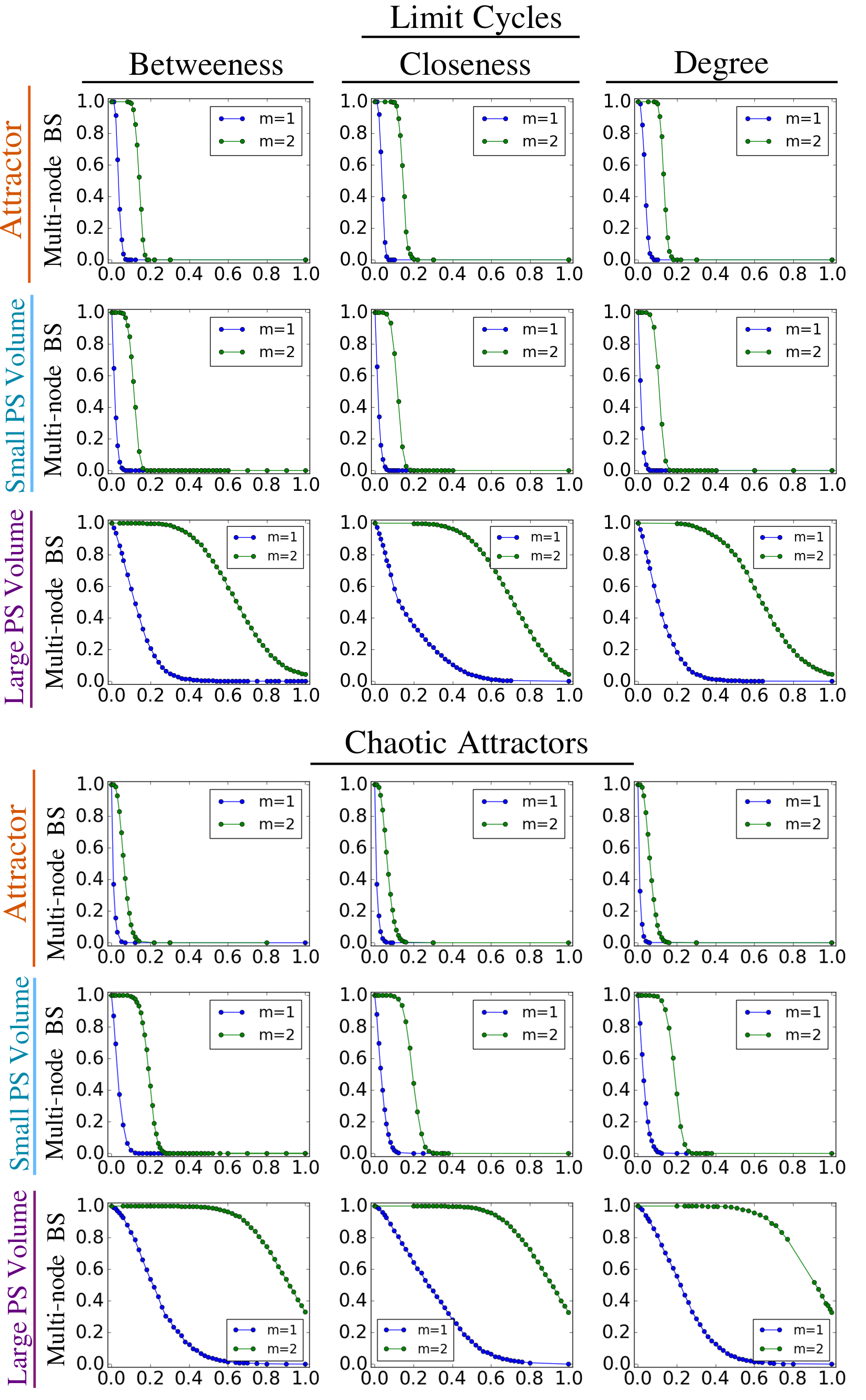}};
	\node[scale=0.6, rotate=0,left of= g] (specJ) at (-3.25,-7.4) {\bf \textcolor{black!70}{\Large \textit{f}}};
	\node[scale=0.6, rotate=0,left of= g] (specJ) at (-0.5,-7.4) {\bf \textcolor{black!80}{\Large \textit{f}}};
	\node[scale=0.6, rotate=0,left of= g] (specJ) at (2.2,-7.4) {\bf \textcolor{black!90}{\Large \textit{f}}};
	\end{tikzpicture}
	\caption{Dependence of the multi-node Basin Stability of scale-free networks of multi-stable oscillators, on the fraction $f$ of perturbed nodes, with $a=0.1$ (multi-stable limit cycles) and $a=0.36$ (multi-stable chaotic attractors). Here $N=100$, $C=1$, and $m=1$ and $m=2$. The perturbed nodes are chosen in descending order of (left panel) betweeness centrality, (middle panel) closeness centrality and (right panel) degree, i.e. the perturbed nodes are the ones with the highest $b$, $c$ or $k$. Attractor switching, small phase-space volume and large phase-space volume perturbation schemes are considered (cf. Fig.~\ref{main_fig}). The trends are almost invariant for different $N$.
        }
	\label{comapre_m1m2}
\end{figure}  


Another noticeable trend is the following: the scale-free network is significantly more robust under perturbations that push the perturbed nodes to states inside a large volume of phase space, rather than perturbations that take the perturbed nodes to a state inside a small phase-space volume in the basin of attraction of the competing attractor, or causes switching of attractors. This can be rationalized as follows: perturbations onto a large phase-space volume, while large in magnitude, do not necessarily push the perturbed node onto the basin of the competing attractor, as this phase-space volume includes basins of both attractors (cf. Fig.~\ref{main_fig}). So now we have a lower probability of the perturbed node migrating into the basin of the other attractor, and hence the network on an average needs a larger fraction of nodes to de-stabilize the collective dynamics (i.e. $f_{crit}$ is larger).

Fig. \ref{fnode_properties} also shows the Basin Stability of a network where the perturbed nodes are randomly chosen, corresponding to {\em random attacks} on a subset of nodes. Clearly, a {\em targetted attack on nodes with high centrality can destroy the collective dynamics much more efficiently than random attacks, irrespective of the nature of the nodal dynamics and type of perturbation.} 

Another important feature that can be inferred from Fig. \ref{fnode_properties} is the critical fraction of perturbed nodes that lead to loss of recovery for networks with different types of dynamical attractors. We find that networks of oscillators with multi-stable limit cycles are more robust than those with multi-stable chaotic attractors under the attractor switching. However, counter-intuitively networks of multi-stable limit cycle oscillators are less robust than networks of multi-stable chaotic attractors, under the perturbations onto small or large phase-space volumes. This can be rationalized by taking into account the more complicated basin boundaries for co-existing chaotic attractors. These complex boundaries make the basins more intertwined and actually help recovery, as the system can re-enter the original basin more readily. So interplay of the nodal dynamics and the type of perturbation plays a significant role in determining the global stability of the collective dynamics.


Finally we estimate the critical fraction of perturbed nodes necessary to destroy the resilience of the scale-free network with $m=1$ and $m=2$ (cf. Fig.~\ref{comapre_m1m2}), focussing on perturbations at nodes of highest centrality. We find that under all three classes of perturbations, in scale-free networks of multi-stable limit cycles, as well as multi-stable chaotic attractors, $f_{crit} \rightarrow 0$ for $m=1$. That is, the smallest non-zero fraction of perturbed nodes of high centrality destroys the ability of the network to recover to its initial state. However, interestingly scale-free networks with $m=2$ are always more robust.
Now, we can use this result, in conjunction with knowledge of the distributions of the different centrality measures in scale-free networks with $m=1$ and $m=2$, to assess which centrality measure of the perturbed node most affects the ability of these networks to recover from perturbations. The important feature of the centrality distributions we can exploit is the following: the tail of the distribution of degree and closeness centrality extends to much lower values for scale-free networks with $m=2$ vis-a-vis networks with $m=1$. So the highest degree and closeness centrality values typically found in networks with $m=1$ are smaller than those found in networks with $m=2$. In contrast, the tail of the distribution of betweeness centrality extends to higher values for the scale-free network with $m=1$ compared to the network with $m=2$, and so typically the highest betweeness values in a network with $m=1$ are larger than that for networks with $m=2$. Additionally note that the set of nodes with the highest degree, closeness and betweeness centrality overlap to a very large extent. This implies that in the scale-free network with $m=2$ these nodes have lower $b$, while having higher $c$ and $k$, than the corresponding set in the network with $m=1$.
So if betweeness centrality of the perturbed nodes is the principal property determining robustness, then the robustness of the network with $m=1$ should be less than the network with $m=2$. However, if degree or closeness centrality of the perturbed nodes is more crucial than betweeness centrality, then networks with $m=2$ should be less robust than those with $m=1$. Since we find that scale-free networks with $m=2$ are always more robust than  $m=1$,
we can infer that the {\em impact of betweeness centrality of the perturbed nodes on the resilience of the collective state is more dominant than the effect of the degree or closeness centrality of the perturbed nodes}. This result has much relevance in dictating which nodes to guard most stringently against localized attacks in order to maintain dynamical robustness of the network.

 

\section{Discussion}

We have explored the collective dynamics of multi-stable Duffing oscillators connected in different network topologies, ranging from rings and star networks to scale-free networks, under diffusive coupling. We estimated the dynamical robustness of such networks by introducing a variant of the concept of multi-node Basin Stability. This measure allowed us to assess the global resilience of the dynamical network in response to large perturbations affecting specific nodes of the system. 

We first considered rings of multi-stable oscillators. Since all nodes have the same centrality in a ring, it provides us with a reference system to understand how the number of perturbed nodes influences resilience. Intersetingly, we find that perturbations at random locations in a ring allows the system to recover its original dynamical more readily than perturbations on a cluster of contiguous nodes. We then went on to study the star network, where the degree, closeness and betweeness centrality of the central hub node is very different from that of the peripheral nodes. This network then provides a good test-bed to investigate the correlation between the centrality of a node and the resilience of the network under large localized perturbations at such nodes. Remarkably we find that perturbing just the hub node is enough to destroy the ability of the network to recover its initial state, irrespective of the nature of the nodal dynamics. On the other hand, the network remains robust under a large perturbations to a very large fraction of the peripheral nodes, i.e. the hub node is capable of steering the vast majority of perturbed peripheral nodes back to their original state. This clearly demonstrates that the centrality of the perturbed node(s) crucially determines the resilience of the network. Also, it is clearly evident that a ring and star network of multi-stable limit cycles is more robust than that comprised of multi-stable chaotic attractors.

Lastly we considered scale-free networks, which are wide-spread in the natural world and in engineered systems. From the conceptual point of view, scale-free networks being heterogeneous enables us to answer more subtle questions regarding the correlation of centrality properties and the potential recovery of the network. First, it is clearly evident again that perturbing nodes with the highest centrality results in extreme reduction of Basin Stability, while the Basin Stability falls slowly when nodes of low centrality are perturbed.
The second important observation is that targetted attacks on nodes with high centrality can destroy the collective dynamics much more efficiently than random attacks, irrespective of the nature of the nodal dynamics and type of perturbation. We also find that the interplay of the nodal dynamics and the type of perturbation plays a significant role in determining the global stability of the collective dynamics in heterogeneous networks of multi-stable systems. For instance, counter-intuitively scale-free networks of multi-stable limit cycles are less robust than networks of multi-stable chaotic attractors when nodes of lowest centrality are perturbed and the perturbation takes the state of the node to a random phase point in a small or large volume of phase-space.
Finally, we demonstrate that scale-free networks with $m=2$ are more robust than networks with $m=1$. This allows us to infer that the impact of betweeness centrality of the perturbed nodes on the resilience of the collective state is more dominant than the effect of the degree or closeness centrality of the perturbed nodes.

In summary, targetted attack on nodes with high centrality in a star and scale-free networks, and targetted attacks on clusters of nodes in a ring, can destroy the collective dynamics much more efficiently than random attacks. This feature appears to be general and holds for networks of coupled multi-stable limit cycles, as well as coupled multi-stable chaotic attractors.
So these results can serve as a guide to identify the nodes to guard most stringently against targetted attacks in order to maintain dynamical robustness of the network.

\section{Acknowledgements}

SS would like to acknowledge support from the J.C. Bose National Fellowship (SB/S2/JCB-013/2015).
CM would like to acknowledge the financial support from DST INSPIRE Fellowship, India.

\end{document}